\documentclass[twocolumn]{aastex61}

\usepackage{xspace}
\usepackage{xcolor}
\usepackage{amsmath}
\usepackage{amssymb}

\received{November 30, 2017}
\revised{February 6, 2018}
\accepted{February 12, 2018}
\submitjournal{ApJ}

\shorttitle{Future Constraints from FRBs}
\shortauthors{Walters et al.}

 \newcommand{\er}[1]{(\ref{#1})}          	
 
\newcommand{\dmhg}{{\rm DM}_{\mathrm{HG}} }
\newcommand{\dmhgl}{{\rm DM}_{\mathrm{HG,loc}} }
\newcommand{\shgl}{\sigma_{\mathrm{HG,loc}} }
\newcommand{\shgli}{\sigma_{\mathrm{HG,loc},i} }
\newcommand{\dmigm}{{\rm DM}_{\mathrm{IGM}} }
\newcommand{\dmmw}{{\rm DM}_{\mathrm{MW}} }
\newcommand{\dmobs}{{\rm DM}_{\mathrm{obs}} }
\newcommand{\dme}{{\rm DM}_{\mathrm{E}} }
\newcommand{\dmei}{{\rm DM}_{\mathrm{E},i} }
\newcommand{\figm}{f_{\mathrm{IGM}} }

\newcommand{\omde}{\Omega_{\mathrm{DE}} }
\newcommand{\zlim}{z_{\mathrm{lim}}}
\newcommand{\sigm}{\sigma_{\mathrm{IGM}}}
\newcommand{\sigmi}{\sigma_{\mathrm{IGM},i } }

\newcommand{\nfrb}{N_{\mathrm{FRB}}}

\newcommand{\pccm}{{\rm pc}\,{\rm cm}$^{-3}$\xspace}
\newcommand{\lcdm}{$\Lambda$CDM\xspace}
\newcommand{\tfid}{\theta_{\mathrm{fiducial}}}

\newcommand{\obhs}{$\Omega_{\rm b} h^2$\xspace}

 \renewcommand{\L}{Lema\^{\i}tre}
 
\begin{document}

\title{Future Cosmological Constraints from Fast Radio Bursts}

\correspondingauthor{Anthony Walters}
\email{tony.walters@uct.ac.za}

\author[0000-0003-1766-9846]{Anthony Walters}
\affiliation{Department of Mathematics and Applied Mathematics, University of Cape Town, Cape Town, South Africa}
\affiliation{School of Chemistry and Physics, University of KwaZulu-Natal, Durban, 4000, South Africa}

\author{Amanda Weltman}
\affiliation{Department of Mathematics and Applied Mathematics, University of Cape Town, Cape Town, South Africa}

\author{B. M. Gaensler}
\affiliation{Dunlap Institute for Astronomy and Astrophysics, University of Toronto, Toronto, ON M5S 3H4, Canada}

\author{Yin-Zhe Ma}
\affiliation{School of Chemistry and Physics, University of KwaZulu-Natal, Durban, 4000, South Africa}
\affiliation{NAOC--UKZN Computational Astrophysics Centre (NUCAC), University of KwaZulu-Natal, Durban, 4000, South Africa}

\author{Amadeus Witzemann}
\affiliation{Department of Physics and Astronomy, University of the Western Cape, Cape Town, South Africa}
\affiliation{Department of Mathematics and Applied Mathematics, University of Cape Town, Cape Town, South Africa}



\begin{abstract}
We consider the possible observation of Fast Radio Bursts (FRBs) with planned future radio telescopes, and investigate how well the dispersions and redshifts of these signals might constrain cosmological parameters. We construct mock catalogues of FRB dispersion measure (DM) data and employ Markov Chain Monte Carlo (MCMC) analysis, with which we forecast and compare with existing constraints in the flat \lcdm model, as well as some popular extensions that include dark energy equation of state and curvature parameters. We find that the scatter in DM observations caused by inhomogeneities in the intergalactic medium (IGM) poses a big challenge to the utility of FRBs as a cosmic probe. Only in the most optimistic case, with a high number of events and low IGM variance, do FRBs aid in improving current constraints. In particular, when FRBs are combined with CMB+BAO+SNe+$H_0$ data, we find the biggest improvement comes in the \obhs constraint. Also, we find that the dark energy equation of state is poorly constrained, while the constraint on the curvature parameter $\Omega_k$, shows some improvement when combined with current constraints. When FRBs are combined with future BAO data from 21cm Intensity Mapping (IM), we find little improvement over the constraints from BAOs alone. However, the inclusion of FRBs introduces an additional parameter constraint, \obhs, which turns out to be comparable to existing constraints. This suggest that FRBs provide valuable information about the cosmological baryon density in the intermediate redshift Universe, independent of high redshift CMB data.
\end{abstract}




\section{Introduction}
Improvements in cosmological measurement in recent years have been said to hail an era of ``precision cosmology'', with observations of the cosmic microwave background (CMB) temperature anisotropies \citep{2013ApJS..208...19H, 2016A&A...594A..13P, 2016A&A...594A..14P}, baryon acoustic oscillation (BAO) wiggles in the galaxy power spectrum  \citep{2011MNRAS.416.3017B, 2014MNRAS.441...24A, 2015MNRAS.449..835R}, luminosity distance-redshift relation of Type~Ia supernovae (SNIa) \citep{2004ApJ...607..665R, 2007ApJ...659...98R, 2008ApJ...686..749K, 2014A&A...568A..22B}, local distance ladder \citep{2016ApJ...826...56R}, galaxy clustering and weak lensing \citep{2017arXiv170801530D}, and direct detection of gravitational waves \citep{2017Natur.551...85A}, providing constraints on cosmological model parameters at percent, or sub-percent, level precision. Since the discovery of the accelerated expansion of the Universe, these observations have cemented the emergence of the flat $\Lambda$CDM model as the standard model of cosmology, in which global spatial curvature is zero, and the energy budget of the Universe is dominated by  ``dark energy'' in the form of a cosmological constant, $\Lambda$. However, beyond the \lcdm paradigm there are a large number of dark energy models aimed at explaining the accelerated expansion of the Universe (see reviews \citep{2011CoTPh..56..525L, 2015PhR...568....1J}, and references therein), and so understanding the nature of dark energy remains one of the central pursuits in modern cosmology. To this end, it has become common observational practice to constrain the dark energy equation of state, $w(z)$, and check for deviations from the \lcdm value of $w=\mathrm{const.}=-1$. While observational probes do not indicate any significant departure from  $\Lambda$CDM \citep{2017arXiv170901091H},  there is still room to tighten constraints and thereby rule out competing alternatives for dark energy. In particular, by tuning the parameters of alternative theories of dark energy, one can recover the behaviour of $\Lambda$CDM model at both the background expansion and perturbation levels~\citep{2011CoTPh..56..525L,2015PhR...568....1J}. 

Observations of the CMB together with SNIa and BAO constrain the spatial curvature parameter to be very small, $|\Omega_k|<0.005$ \citep{2016A&A...594A..13P}, consistent with the flat \lcdm model, and the inflationary picture of the early Universe. However, model independent constraints from low redshift probes are not nearly as strong, with SNIa alone preferring an open universe with $\Omega_k \sim 0.2$ \citep{2015PhRvL.115j1301R}.   Similarly, constraints on the baryon fraction, $\Omega_{\mathrm b}$,  derived from observations of the CMB, and the abundance of  light elements  together with the theory of Big Bang Nucleosynthesis (BBN) \citep{2016ApJ...830..148C}, are both rooted in high redshift physics. And while these constraints are somewhat consistent, the BBN results strongly depend on nuclear cross section data \citep{2016ApJ...830..148C,2016MNRAS.458L.104D}. Thus, independent and precise low redshift probes of spatial curvature and the baryon density parameter which confirm the constraints from high redshift data are of observational and theoretical interest.

Recently, a promising new astrophysical phenomenon, so called Fast Radio Bursts (FRBs) \citep{2007Sci...318..777L, 2011MNRAS.415.3065K, 2013Sci...341...53T, 2014ApJ...790..101S, 2015MNRAS.447..246P, 2014ApJ...792...19B, 2015ApJ...799L...5R, 2016MNRAS.460L..30C, 2015Natur.528..523M, 2016Natur.530..453K, 2016Sci...354.1249R, 2017MNRAS.468.3746C, 2017MNRAS.469.4465P}, has emerged. An FRB is characterised by a brief pulse in the radio spectrum with a large dispersion in the arrival time of its frequency components, consistent with the propagation of an electromagnetic wave through a cold plasma. To date a total of 25 such FRBs \footnote{From version 2.0 of the FRB catalogue \citep{2016PASA...33...45P} found at {\tt http://www.frbcat.org/}, accessed on 17 November 2017 } have been detected, primarily by the the Parkes Telescope in Australia, but more recently interferometric detections have also been reported. Considering the greatly improved sensitivity of upcoming radio telescopes, expectations are high that many more FRB events will be observed in the near future \citep{2017MNRAS.465.2286R, 2017ApJ...846L..27F}. While their exact location and formation mechanism is still a subject of ongoing research \citep{2013ApJ...776L..39K, 2013PASJ...65L..12T, 2014ApJ...780L..21Z, 2015MNRAS.450L..71F, 2014MNRAS.442L...9L,  2016MNRAS.457..232C, 2017MNRAS.465L..30G, 2016ApJ...823L..28G, 2016ApJ...822L...7W, 2017ApJ...843L..26B, 2017arXiv170806352L, 2017MNRAS.468.2726K, 2017MNRAS.469L..39K, 2017arXiv170807507G, 2017ApJ...844..162T}, their excessively large dispersion measures (DMs) argue that they have an extragalactic origin \citep{2015RAA....15.1629X}. Indeed, one FRB event has been sufficiently localised to be associated with a host galaxy at $z=0.19$ \citep{2017ApJ...834L...7T}. Should one be able to associate a redshift with enough FRBs, it would give access to the $\mathrm{DM}(z)$ relation, which may provide a new probe of the cosmos \citep{2014ApJ...783L..35D, 2014PhRvD..89j7303Z,  2014ApJ...788..189G, 2016ApJ...830L..31Y, 2017A&A...606A...3Y}, possibly complementary to existing techniques. In addition, the observation of strongly lensed FRBs may help to constrain the Hubble parameter \citep{2017arXiv170806357L} and the nature of dark matter \citep{2016PhRvL.117i1301M}, and dispersion space distortions may provide information on matter clustering \citep{2015PhRvL.115l1301M}, all without redshift information.

In this paper we assess the potential for using FRB  $DM(z)$ measurements, to constrain the parameter space of various cosmological models, and whether this may improve the existing constraints coming from other observations. The outline is as follows: The details of modelling an extragalactic population of FRBs, constructing a mock catalogue of DM observations, and extracting and combining cosmological parameter constraints is given in \S\ref{cosmoFRB}. Parameter constraint forecasts from the mock FRB data, and its combination with CMB + BAO + SNIa + $H_0$ (hereafter referred to as CBSH), is given in \S\ref{base} for the flat \lcdm model, and in \S\ref{ext} for 1- and 2-parameter extensions to the flat \lcdm model. Possible synergies with other experiments are discussed in \S\ref{synergies}.

\section{Cosmology with Fast Radio Bursts}
\label{cosmoFRB}
\subsection{Dispersion of the Intergalactic Medium}
\label{cosmoDM}
The DM of an FRB is associated with the propagation of a radio wave through a cold plasma, and is related to the path length from the emission event to observation, and the distribution of free electrons along that path, ${\rm DM}=\int n_{\rm e} {\rm d}l$. If FRBs are of extragalactic origin their observed dispersion measure, $\dmobs$, should be the sum of a number of different contributions, namely; from propagating through its host galaxy, $\dmhg$, the intergalactic medium (IGM), $\dmigm$, and the Milky Way, $\dmmw$ \citep{2014ApJ...783L..35D}.  Since $\dmmw$ as a function of Galactic latitude  is well known from pulsar observations \citep{2017ApJ...835...29Y}, and its contribution  to $\dmobs$ is relatively small in most cases, we assume it can be reliably subtracted. We choose to work with the extragalactic dispersion measure, given by \citep{2016ApJ...830L..31Y}
\begin{align}
\dme \equiv \dmobs - \dmmw = \dmigm + \dmhg, \label{dme}
\end{align}
where $\dmhg$ is defined in the observers frame, and related to that at the emission event by
\begin{align}
\dmhg = \frac{\dmhgl}{1+z}.
\end{align}
This contribution is not well known and is expected to depend on the type of host galaxy, its inclination relative to the observer, and the location of the FRB inside the host galaxy \citep{2015RAA....15.1629X, 2016ApJ...830L..31Y}, and so we include this as a source of uncertainty in our analysis.

The intergalactic medium is inhomogeneous and so $\dmigm(z)$ will have a large sightline-to-sightline variance, with estimates ranging between $\sim200$ and 400~pc cm$^{-3}$ by $z\sim 1.5$ \citep{2014ApJ...780L..33M}. It has however been shown that with enough FRB events in small enough redshift bins, the mean dispersion measure in each bin will approach the Friedmann-\L-Robertson-Walker (FLRW) background value to good approximation. Specifically, with $N\sim80$ events in the redshift bin $1\leq z \leq 1.05$, the mean dispersion measure will be with 5\% of the FLRW background value, at 95.4\% confidence \citep{2014PhRvD..89j7303Z}. This is essential if one wishes to measure the cosmological parameters with any precision.

Assuming a non-flat FLRW Universe that is dominated by matter and dark energy,  one finds the average (background) dispersion measure of the intergalactic medium is \citep{2014ApJ...783L..35D, 2014PhRvD..89j7303Z, 2014ApJ...788..189G}
\begin{align}
\langle \dmigm(z) \rangle =  \frac{3 c H_0 \Omega_{\rm b} \figm }{8 \pi G m_{\rm p}} \int^z_0 \frac{\chi (z')  (1+z') }{E(z')}~ {\rm d}z',\label{dmigm}
\end{align}
where
\begin{align}
E(z)&=\left[ (1+z)^3 \Omega_{\rm m} + f(z) \Omega_{\rm DE} + (1+z)^2 \Omega_k \right]^{1/2},\\
\chi(z) &= Y_{\rm H} \chi_{\rm e,H}(z) + \frac{1}{2}Y_{\rm p} \chi_{\rm e,He}(z), \label{chiz}\\
f(z) &= \exp\left[ 3 \int_0^z \frac{(1 + w(z'')){\rm d}z''}{(1+z'')}\right],  \label{fz}
\end{align}
and $H_0$ is the value of the Hubble parameter today, $\Omega_{\rm b}$ is the baryon mass fraction of the Universe, $\figm$ is the fraction of baryon mass in the intergalactic medium, $Y_{\rm H}=3/4$  ($Y_{\rm p}=1/4$) is the hydrogen  (helium) mass fraction in the intergalactic medium, and $\chi_{\rm e,H}$ ($\chi_{\rm e,He}$) is the ionisation fraction of hydrogen (helium).  The cosmological density parameters for matter and curvature are $\Omega_{\rm m}$ and $\Omega_k$, respectively, and the dark energy density parameter is given by the constraint  $\omde \equiv 1-\Omega_{\rm m}-\Omega_k$. 

We allow for the equation of state of dark energy, $w$, to vary with time, and parameterise it by \citep{2001IJMPD..10..213C, 2003PhRvL..90i1301L}
\begin{align}
w(z) &= w_0 +w_a \frac{z}{1+z} \label{cpl},
\end{align}
where $w_0$ and $w_a$ are the CPL parameters. Substituting \er{cpl} into \er{fz}, and integrating, gives an exact analytic expression for the growth of dark energy density as a function of redshift
\begin{align}
f(z)=(1+z)^{3(1+w_0+w_a)} \exp{\left[-3 w_a\frac{ z}{1+z}\right]}.
\label{fzCPL}
\end{align}
Choosing $(w_0,w_a)=(-1,0)$ in \er{fz} gives $f(z)=\mathrm{const.}$, corresponding to the $\Lambda$CDM model, in which dark energy is a cosmological constant.

For simplicity (to avoid modelling any astrophysics) we restrict our analysis to the region $z\leq3$, since current observations suggest that both hydrogen and helium are fully ionised there \citep{2009RvMP...81.1405M, 2011MNRAS.410.1096B}, and thus we can safely take $\chi_{e,H}=\chi_{e,He}=1$ in \er{chiz}. This gives a constant $\chi(z)=7/8$ in the region of interest. The $\figm$ term presents some complications. Strictly speaking, $\figm$ is a function of redshift ($\figm=\figm(z)$) ranging from about 0.9 at $z\gtrsim1.5$ to $0.82$ at $z \leq 0.4$ \citep{2009RvMP...81.1405M, 2012ApJ...759...23S}, and should be included inside the integral in \er{dmigm}. As a first approximation we neglect the effect of evolving $\figm$, and set it to a constant.

\subsection{Telescope Time and the Mock Catalogue}
\label{forecast}
Based on current detections,  the FRB event rate in the Universe is expected to be high, and given the improved design sensitivity of future radio telescopes, their detection rate is expected to increase significantly. This value, of course, will depend on the exact specifications of the telescope, and the true distribution and spectral profile of FRBs.  For example, assuming they live only in low mass host galaxies, and have a Gaussian-like spectral profile, the mid-frequency component of the Square Kilometre Array (SKA) is expected to detect FRBs out to $z\sim3.2$ at a rate of $\sim10^3$ sky$^{-1}$ day$^{-1}$ \citep{2017ApJ...846L..27F}. In the more immediate future, the Hydrogen Intensity Real-time Analysis eXperiment (HIRAX) \citep{2016SPIE.9906E..5XN} and the Canadian Hydrogen Intensity Mapping Experiment (CHIME) \citep{2014SPIE.9145E..22B}, are expected to detect $\sim 50-100$ day$^{-1}$ and $\sim 30-100$ day$^{-1}$, respectively \citep{2017MNRAS.465.2286R}.  Assuming that 5\% of the detected FRBs can be sufficiently localised to be associated with a host galaxy, the rate of detection and localisation would be roughly $\sim2-5$ day$^{-1}$ for HIRAX and CHIME, and far higher for the SKA.  This suggests that a large catalogue of localised FRBs could be built up relatively quickly, and the main bottleneck in obtaining a catalogue of $DM(z)$ data will be acquiring the redshifts. Given the bright emission lines in the spectrum of the host galaxy for the repeating FRB~121102 \citep{2017ApJ...834L...7T}, a mid- to large-sized optical telescope should be able to obtain $\sim10$ redshifts for FRB host galaxies per night; we thus estimate that a redshift catalogue with $\nfrb=1000$ will take approximately 100 nights of observing to construct, which would be feasible with a dedicated observing program spread over a few years.  

Motivated by a phenomenological model for the distribution of gamma ray bursts,  we assume the redshift distribution of FRBs is given by $P(z)=z e^{-z}$ \citep{2014PhRvD..89j7303Z, 2016ApJ...830L..31Y}, and simulate $\dme(z)$ measurements, given by the far right side of \er{dme}.  Due to matter inhomogeneities in the IGM, and variations in the properties of the host galaxy, we promote $\dmigm$ and $\dmhgl$ to random variables, and sample them from a normal distribution. That is $\dmigm \sim  \mathcal{N} \left( \langle \dmigm (z)\rangle, \sigm \right)$, and $\dmhgl \sim \mathcal{N} \left( \langle \dmhgl \rangle, \shgl \right)$. We assume  $\langle \dmigm (z)\rangle$ is given by \er{dmigm} and a flat $\Lambda$CDM background as the fiducial cosmology, using the best fit CBSH parameter values provided by the Planck 2015 data release\footnote{Planck 2015 covariance matrices and MCMC chains can be found at \tt http://pla.esac.esa.int/pla/\#cosmology}, listed in the second column of table \ref{multi_table}. We also take  $\figm=0.83$ \citep{2012ApJ...759...23S}. 


The value of $\dmhgl$ is expected to contain contributions from the Interstellar Medium (ISM) of the FRB host galaxy and near-source plasma. Since FRB progenitors and their emission mechanisms are as yet unknown, reasonable values of  $\langle\dmhgl \rangle$ and $\shgl$ are still debatable. Here we assume nothing about the host galaxy type or location of the FRB therein, just that there is a significant contribution to $\dmhgl$ due to near source-plasma, and thus take $ \langle \dmhgl \rangle=200$ pc cm$^{-3}$ and $\shgl=50$ pc cm$^{-3}$ \citep{2016ApJ...830L..31Y}.  To investigate the effect of sample size and IGM inhomogeneities  on resulting constraints, we construct a number of mock catalogues with various values for $\sigm$ and $\nfrb$. For the most optimistic sample, we choose $(\nfrb,\sigm)=(1000,200)$. See table \ref{cats} for a summary of the various catalogues.
\begin{table}[!htb]
\centering
\begin{tabular} { l c c c }
 & $\nfrb$  & $\sigm$ [\pccm] & $\zlim$ \\
\hline
FRB1  & $1000$		& $200$	& $3$\\
FRB2  & $1000$		& $400$& $3$	\\
FRB3  & $100$		& $200$& $3$	\\
\hline
\end{tabular}
\caption{Parameter values used when populating the various mock FRB catalogues. The number of FRB events is shown in the first column, the sightline-to-sightline variance in the second column, and the limiting redshift in the third column.}
\label{cats}
\end{table}

\subsection{Parameter Estimation and Priors}
For the MCMC analysis we use the $\chi^2$ statistic as a measure of likelihood for the parameter values. The log-likelihood function is given by
 \begin{align}
 \ln \mathcal{L_{\mathrm{FRB}}}(\theta|d) = -\frac{1}{2} \sum_i \frac{\left( \dmei - \langle \dme \rangle  \right)^2}{ \sigmi^2   + \left[ \shgli /(1+z_i)\right]^2} , \label{likelihood}
 \end{align}
where $\theta$ is the set of fitting parameters, $d$ is the FRB data, and the sum over $i$ represents the sequence of FRB data in the sample. Constraints on the flat $\Lambda$CDM model parameters are obtained by setting $\Omega_k=0$ in \er{dmigm} and $w=-1$ in \er{fz}, and then fitting the mock data for $\theta=(\Omega_{\rm m}, H_0, \Omega_{\rm b} h^2, \langle\dmhgl\rangle )$.  To investigate spatial curvature in the \lcdm model, we allow for $\Omega_k\neq0$ in \er{dmigm}, and include it as an additional fitting parameter. For the dark energy constraints we consider two model parametrisations with flat spatial geometry. In the first case, we extend  to the $w$CDM model, allowing for $w=\mathrm{const.}\neq-1$. We set $\Omega_k=0$ in \er{dmigm} and $(w_0,w_a)=(w,0)$ in \er{cpl}, and fit the data for $\theta=(w, \Omega_{\rm m}, H_0, \Omega_{\rm b} h^2 , \langle \dmhgl \rangle)$. In the second case, we allow for dark energy to vary with time and use the CPL parametrisation \er{cpl}, and thus set $\Omega_k=0$ in \er{dmigm}, and fit the FRB data for the parameters $\theta=(w_0, w_a, \Omega_{\rm m}, H_0, \Omega_{\rm b} h^2 , \langle \dmhgl \rangle)$. For all the extended models, we fit to the flat \lcdm data described in \S\ref{forecast}, and examine how close to fiducial values the additional parameters are constrained. This also allows us to easily combine the constraints with existing data, which is consistent with flat \lcdm.

We use the Python package {\it emcee} \citep{2013PASP..125..306F} to determine the posterior distribution for the parameters, and {\it GetDist}\footnote{Package available at {\tt https://github.com/cmbant/getdist}} for plotting and analysis. When prior information is included in the analysis we use the respective covariance matrix provided by the Planck 2015 data release. We thus calculate the priors according to 
\begin{align}
\ln P(\theta)  = -\frac{1}{2} \xi \mathbf{C}^{-1} \xi,
\label{lnP}
\end{align}
where $P(\theta)$ is the prior probability associated with the parameter values $\theta$, $\mathbf{C}$ is a (square) covariance matrix,  and $\xi = \theta-\tfid$ is the displacement  in parameters space between the relevant parameter values  and the fiducial values. To avoid rescaling the CBSH covariance matrix to accommodate for $\Omega_{\rm b}$, we set up our code to fit for \obhs, which is a primary parameter in the Planck analysis, and thus its covariance is provided.

\begin{figure*}[!ht]
\centering
\includegraphics[width=\linewidth]{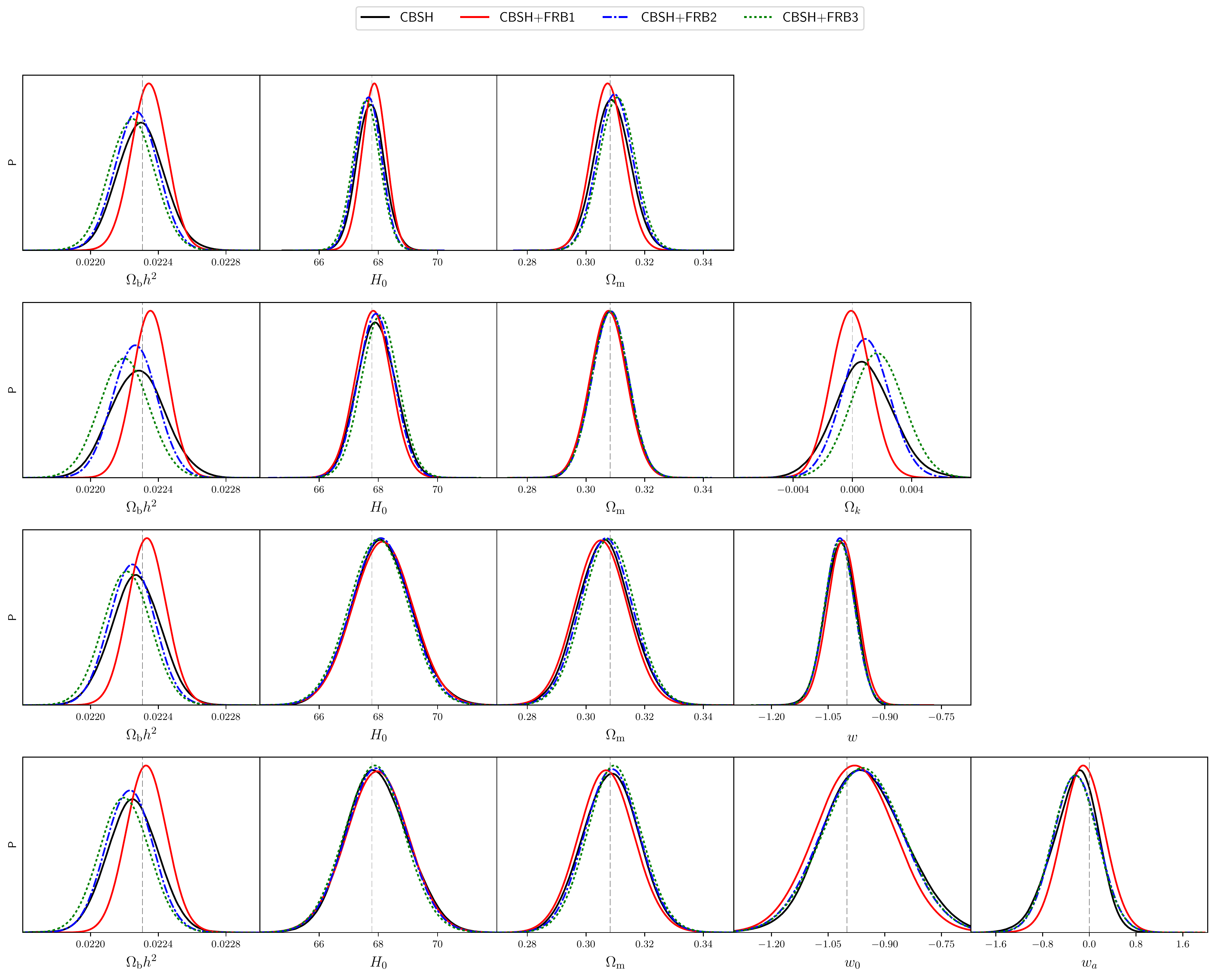}
\caption{Marginalised posterior probability distributions obtained from a combination of CBSH constraints and the various mock FRB catalogues listed in Table \ref{cats}, for all cosmological model parametrisations considered here. From the top to bottom row we list constraints for flat $\Lambda$CDM,  \lcdm with spatial curvature, flat $w$CDM, and flat $w_0w_a$CDM. Black lines indicate the CBSH constraints used as priors. The red, blue and green lines indicate the constraints when CBSH is combined with the FRB1, FRB2 and FRB3 catalogues, respectively. Dashed grey lines indicates the true parameter values used in the mock FRB catalogues (corresponding to the CBSH best fit parameter values in the flat \lcdm model). The area under all curves has been normalised to unity.}
\label{multi_1d}
\end{figure*}

\begin{table*}
\centering
\begin{tabular} { l c c c c }
 Parameter &  95\% limits\\
&CBSH & CBSH+FRB1 & CBSH+FRB2 & CBSH+FRB3\\
 \hline
{\boldmath$ 10~\Omega_{\rm m}       $}& $3.09^{+0.12}_{-0.12}   $& $3.07^{+0.11}_{-0.11}   $ &$3.10^{+0.12}_{-0.12}   $ & $3.11^{+0.12}_{-0.12}   $\\

{\boldmath$H_0                  $}  & $67.74^{+0.92}_{-0.90}     $ &$67.86^{+0.79}_{-0.80}     $ &$67.66^{+0.86}_{-0.87}     $& $67.60^{+0.89}_{-0.89}     $\\

{\boldmath$10^2 ~\Omega_{\rm b} h^2 $}&$2.230^{+0.027}_{-0.026}$ &  $2.235^{+0.021}_{-0.021}$ &$2.227^{+0.025}_{-0.025}$& $2.224^{+0.026}_{-0.026}$\\

{\boldmath$\langle \dmhgl \rangle $}    & &$215^{+30}_{-30}           $& $189^{+60}_{-60}           $&$161^{+90}_{-90}           $\\
 
 \hline
 {\boldmath$10^3 ~ \Omega_k        $} & $0.8^{+4.0}_{-3.9}$ &$-0.1^{+2.6}_{-2.6}$& $0.9^{+3.2}_{-3.2}$ & $1.7^{+3.5}_{-3.5}$ \\

{\boldmath$10~ \Omega_{\rm m}       $} &$3.08^{+0.12}_{-0.12}   $ & $3.08^{+0.12}_{-0.12}   $ & $3.08^{+0.12}_{-0.12}   $ & $3.08^{+0.12}_{-0.12}   $\\\

{\boldmath$H_0                  $} & $67.9^{+1.3}_{-1.2}        $&$67.8^{+1.2}_{-1.2}        $  & $67.9^{+1.2}_{-1.2}        $ & $68.0^{+1.2}_{-1.2}        $ \\

{\boldmath$10^2 ~ \Omega_{ \rm b} h^2 $} & $2.228^{+0.032}_{-0.031}$ &$2.235^{+0.020}_{-0.021}$ & $2.226^{+0.026}_{-0.026}$ & $2.220^{+0.029}_{-0.029}$\\

{\boldmath$\langle \dmhgl \rangle $} &&$201^{+40}_{-40}           $  & $196^{+60}_{-60}           $ & $177^{+90}_{-90}           $ \\
\hline
 
{\boldmath$w              $} & $-1.019^{+0.075}_{-0.080}  $& $-1.012^{+0.077}_{-0.078}  $&  $-1.020^{+0.077}_{-0.077}  $  &$-1.020^{+0.077}_{-0.077}  $\\

{\boldmath$10~ \Omega_{\rm m}       $} & $3.06^{+0.18}_{-0.18}   $& $3.05^{+0.18}_{-0.17}   $ & $3.07^{+0.17}_{-0.17}   $& $3.08^{+0.17}_{-0.17}   $\\

{\boldmath$H_0                  $}  & $68.1^{+2.1}_{-1.9}        $ & $68.1^{+2.0}_{-2.0}        $   & $68.1^{+1.9}_{-1.9}        $    & $68.0^{+1.9}_{-1.9}        $          \\

{\boldmath$10^2 ~ \Omega_{\rm b} h^2 $}& $2.227^{+0.027}_{-0.029}$      & $2.233^{+0.022}_{-0.022}$    & $2.224^{+0.026}_{-0.026}$  & $2.221^{+0.028}_{-0.028}$        \\

{\boldmath$\langle \dmhgl \rangle $} && $204^{+30}_{-30}           $& $191^{+60}_{-60}           $  & $163^{+90}_{-90}           $   \\
\hline

{\boldmath$w_0            $} & $-0.95^{+0.21}_{-0.20}     $& $-0.98^{+0.21}_{-0.21}     $& $-0.96^{+0.21}_{-0.21}     $& $-0.96^{+0.21}_{-0.21}     $\\

{\boldmath$w_a            $}& $-0.25^{+0.72}_{-0.78}     $& $-0.10^{+0.71}_{-0.71}     $& $-0.23^{+0.76}_{-0.76}     $& $-0.24^{+0.77}_{-0.76}     $\\

{\boldmath$10~ \Omega_{\rm m}       $} &  $3.08^{+0.19}_{-0.19}   $& $3.07^{+0.19}_{-0.18}   $& $3.09^{+0.18}_{-0.18}   $& $3.10^{+0.18}_{-0.18}   $\\

{\boldmath$H_0            $} & $68.0^{+2.0}_{-2.0}        $& $68.0^{+2.0}_{-2.1}        $& $67.9^{+2.0}_{-2.0}        $& $67.9^{+2.0}_{-2.0}        $\\

{\boldmath$10^2 ~ \Omega_{\rm b} h^2   $} & $2.225^{+0.030}_{-0.029}$& $2.233^{+0.024}_{-0.023}$& $2.223^{+0.027}_{-0.027}$ & $2.220^{+0.029}_{-0.029}$ \\

{\boldmath$\langle \dmhgl \rangle  $} && $205^{+30}_{-30}           $& $195^{+60}_{-60}           $ & $167^{+90}_{-90}           $\\
\hline
\end{tabular}
\caption{Parameter constraints for flat \lcdm and some 1- and 2-parameter extensions, namely; \lcdm with spatial curvature, $w$CDM and  $w_0w_a$CDM. The first column lists the constraints, for each model, obtained from the FRB1 catalogue alone, the second column lists the corresponding CBSH constraints, and the third fourth and fifth columns list the combined constraints from the various catalogues listed in Table~\ref{cats}.}
\label{multi_table}
\end{table*}

\section{Parameter Constraints Forecast}
Here we discuss the FRB constraints forecast for the flat \lcdm model and some simple 1- and 2-parameter extensions. In all models, when fitting the most optimistic catalogue, FRB1, we find that $H_0$  and \obhs are unconstrained when no prior information about the parameters is included. This is unsurprising, since $\dmigm \propto \Omega_{\rm b} H_0$. 
And as a result, the other cosmological parameters are only very weakly constrained, if at all. In all models we find the measurement precision of $\Omega_{\rm m}$ is tens of a percent, hardly good enough to be considered a tool for `precision cosmology' at the sub-percent level. We thus include the CBSH covariance matrix in our analysis in order to determine if FRBs offer any additional constraining power.

In figure \ref{multi_1d} we plot a compilation of  the marginalised 1D posterior probability distributions for the cosmological parameters, obtained from a combination of CBSH constraints and the various mock FRB catalogues listed in Table \ref{cats}. Black lines indicate the CBSH constraints used in the covariance matrix for calculating the priors, given by Eq.~\er{lnP}. The solid red, dot-dashed blue and dotted green lines indicate the constraints when CBSH is combined with the FRB1, FRB2 and FRB3 catalogues, respectively. The corresponding 2-$\sigma$ confidence intervals are listed in table \ref{multi_table}.  We deal with the various cosmological models in turn, below.

\subsection{Flat \lcdm}
\label{base}
\begin{figure}[!ht]
\centering
\includegraphics[width=\linewidth]{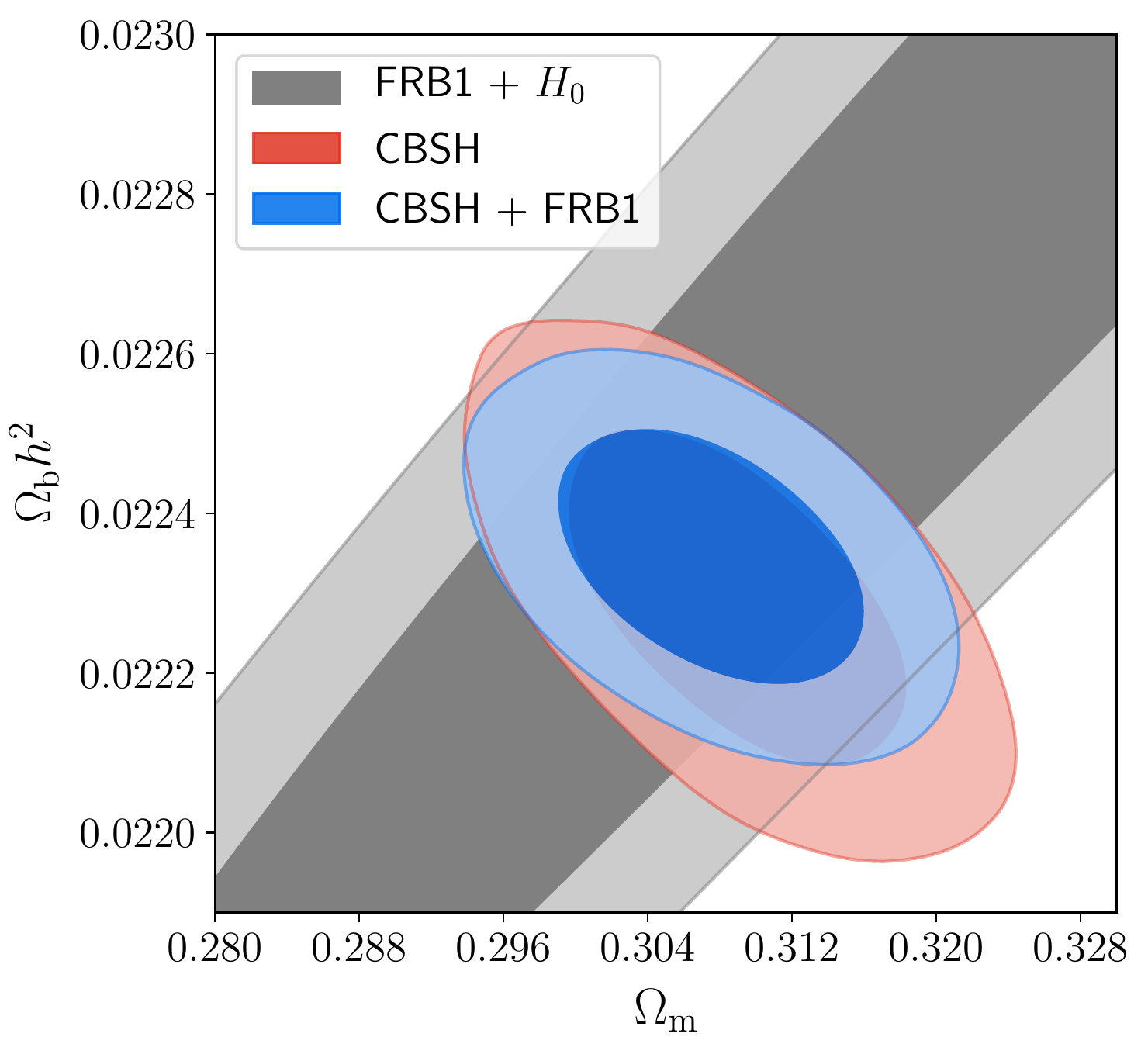}
\caption{ Flat $\Lambda$CDM parameter constraints in the $\Omega_{\rm m}$-$\Omega_{\rm b}h^2$ plane. Constraints obtained from the FRB1 catalogue with a CBSH prior on $H_0$ are shown in grey, the CBSH constraints are shown in red, and the combined constraints are shown in blue. Without including priors, the FRB constraints are very weak, and so have been omitted from this plot. }
\label{base_2Dcomp_omegamomegabh2}
\end{figure}

Including the CBSH covariance matrix gives the combined constraints, CBSH+FRB, shown in the top row of figure \ref{multi_1d}. We find that the posteriors for $H_0$ and $\Omega_{\rm m}$ show only a minor improvement over their priors, as can be seen in the second and third column. The most improved constraint is given by  $\Omega_{\rm b} h^2=0.02235^{+0.00021}_{-0.00021}$, which corresponds to a  $\sim 20\%$ reduction in the size of the  2-$\sigma$ confidence interval of the CBSH prior. The source of this improvement can be seen  in figure \ref{base_2Dcomp_omegamomegabh2}, where we plot constraints in the $\Omega_{\rm m}$-$\Omega_{\rm b}h^2$ plane. Here we include the CBSH prior for $H_0$  with the FRB1 analysis, and plot the resulting constraint (grey) with the CBSH constraints (red). The degeneracy directions of the two eclipses are different, and their intersection gives the combined constrain (blue). Thus, given our current knowledge of the \lcdm parameters and their covariance,  DM observations will provide more information on $\Omega_{\rm b}h^2$ than the other cosmological parameters.

Constraints derived from a combination of CBSH with the various FRB catalogues, represented by the coloured curves in the top row of figure \ref{multi_1d}, illustrate the effect of varying the IGM inhomogeneity and sample size. 
Increasing the IGM inhomogeneity from $\sigm=200$ \pccm to $\sigm=400$ \pccm weakens the constraints considerably. The strongest constraint in this case becomes $\Omega_{\rm b}h^2=0.02227^{+0.00025}_{-0.00025}$, which corresponds to a $\sim 5\%$ reduction in size of the 2-$\sigma$ interval of the CBSH constraint.   Similarly, reducing the samples size to $\nfrb=100$, and keeping IGM inhomogeneity low at $\sigm=200$ \pccm also weakens any improvement offered by FRBs. In this case we find $\Omega_{\rm b}h^2=0.02224^{+0.00026}_{-0.00026}$, which is a $\sim 2\%$ reduction in the size of the CBSH 2-$\sigma$ interval. Clearly one needs many FRB events in order to mitigate the effects of IGM inhomogeneity.

\subsection{Extensions Beyond Flat \lcdm}
\label{ext}
\begin{figure}[!ht]
\centering
\includegraphics[width=\linewidth]{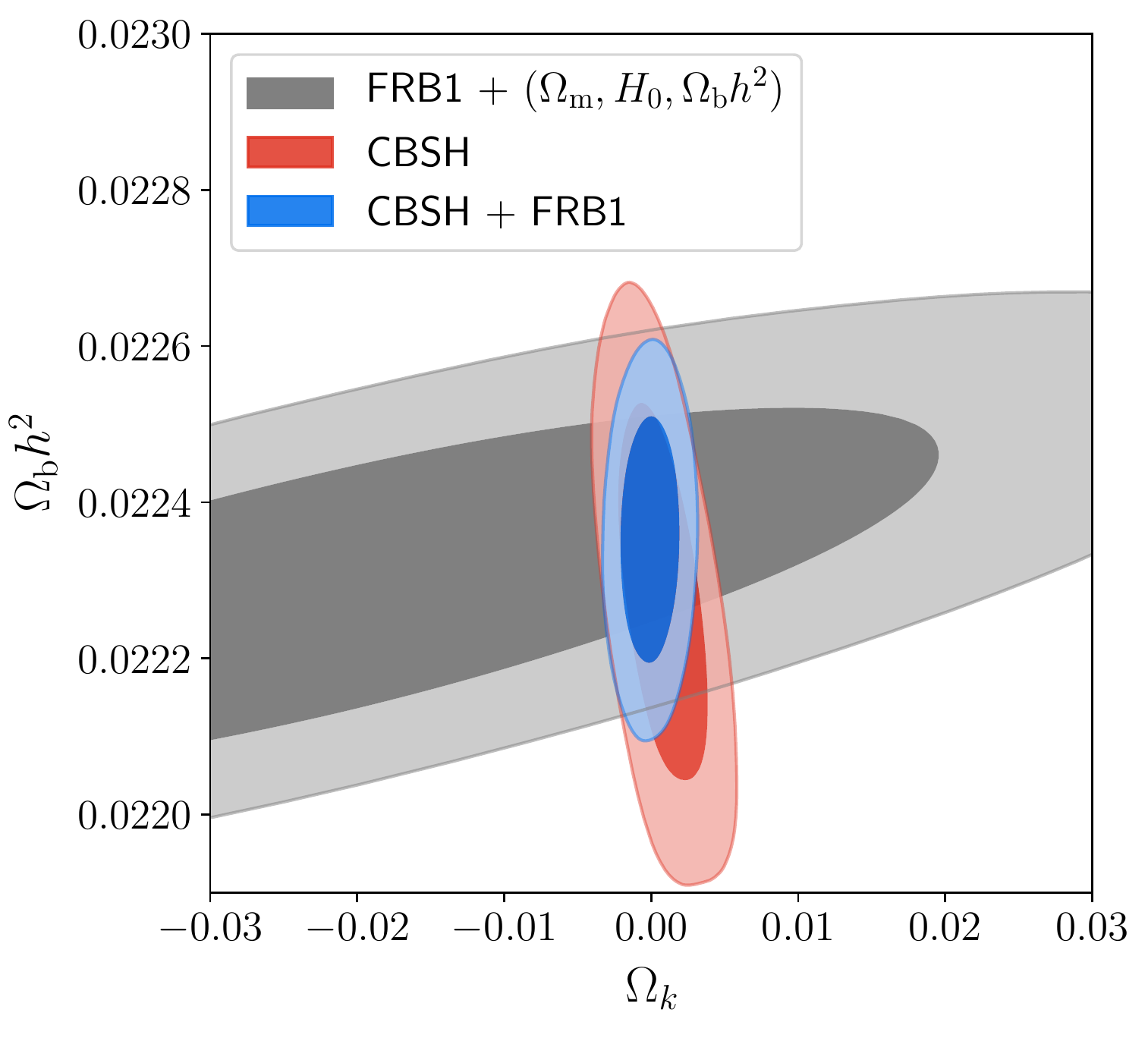}
\caption{Non-Flat \lcdm marginalised 2-D posterior distribution in the $\Omega_{\rm b}h^2$-$\Omega_k$ plane. FRB constraints, when including CBSH covariance for $(\Omega_{\rm m},H_0,\Omega_{\rm b}h^2)$, are shown in grey, CBSH constraints are  shown in red, and the combined constraints are shown in blue. Without including priors, the FRB constraints are very weak, and so have been omitted from this plot.}
\label{base_omegak_2d_omegabh2omegak}
\end{figure}

\paragraph{Curvature}
When no priors are included, we find that $\Omega_k$ is unconstrained by FRB observations alone. Even when the CBSH covariance matrix for  $(\Omega_{\rm m}, H_0, \Omega_{\rm b} h^2 )$ is included, the constraint on $\Omega_k$ remains very weak. However, with the full CBSH covariance matrix included we find $\Omega_k=-0.0001^{+0.0026}_{-0.0026}$ and $\Omega_{\rm b}h^2=0.02235^{+0.00020}_{-0.00021}$. This corresponds to a $\sim 35\%$ reduction in the size of the CBSH 2-$\sigma$ intervals for \obhs and $\Omega_k$. The source of this improvement is illustrated in figure  \ref{base_omegak_2d_omegabh2omegak} where we plot the 2D marginalised constraints in the $\Omega_{\rm b}h^2$-$\Omega_k$ plane. The FRB1 constraints with CBSH covariance for $(\Omega_{\rm m},H_0\Omega_{\rm b}h^2)$ are shown in grey, and the CBSH constraints in red. Its clear that the grey contour very weakly constrains $\Omega_k$. However, it runs orthogonal to the CBSH constraint, and intersects it in a way that simultaneously improves both the $\Omega_k$ and $\Omega_{\rm b} h^2$ constraints when the data are combined, shown in blue.  Posteriors for $\Omega_{\rm m}$ and $H_0$ are dominated by their priors, as can be seen in the second row of figure \ref{multi_1d}.

Increasing the IGM variance to $\sigm=400$ \pccm degrades the constraints to $\Omega_{\rm b}h^2=0.02226^{+0.00026}_{-0.00026}$ and  $\Omega_k=0.0009^{+0.0032}_{-0.0032}$, which corresponds to a $\sim 18\%$ reduction in the size of CBSH 2-$\sigma$ interval. Similarly, reducing the sample size to $\nfrb=100$, we find $\Omega_{\rm b} h^2=0.02220^{+0.00029}_{-0.00029}$ and  $\Omega_k=0.0017^{+0.0035}_{-0.0035}$, which corresponds to a  $\sim10\%$ reduction in the size of the CBSH 2-$\sigma$ intervals. Thus, while FRB observations alone do not constrain $\Omega_k$, they add some constraining power when current parameter covariance is included. As in the flat \lcdm case, many FRBs are needed to realise this improvement.

\paragraph{Testing Concordance}
When the CBSH covariance for $(H_0,\Omega_{\rm b}h^2)$ is included in the analysis, the resulting 2D marginalised  constraint contours are, in all cases, larger than the CBSH ones.  A crucial difference between this result and that of \citep{2014PhRvD..89j7303Z, 2014ApJ...788..189G}, is that the previous authors assumed perfect knowledge of $H_0$ and $\Omega_{\rm b}$, and neglected any contribution from the host galaxy, and thus got a very narrow FRB contour in the $w$-$\Omega_{\rm m}$ plane, which they showed would intersect with, and improve, the current constraints. Alas, we find this is not the case if realistic prior knowledge about $H_0$ and $\Omega_{\rm b}h^2$ is included.

In the third row of figure \ref{multi_1d} we plot the normalised 1D posterior distributions for the $w$CDM model parameters. For all catalogues listed in table \ref{cats} we find that the posteriors are dominated by their priors, with the exception being  $\Omega_{\rm b}h^2$. When using the most optimistic catalogue, we find $\Omega_{\rm b}h^2=0.02233^{+0.00022}_{-0.00022}$, which corresponds to a   $\sim 20\%$ reduction in the size of the 2-$\sigma$ confidence interval of the CBSH prior. Increasing the IGM variance to $\sigm=400$ \pccm weakens this improvement to a few percent. There is no improvement in the \obhs constraint if the sample size is reduced to $\nfrb=100$.

\paragraph{Dynamical Dark Energy}
The normalised 1D posterior distributions can be seen in the bottom row of figure \ref{multi_1d}. With the CBSH covariance included in the FRB1 analysis, we find that all posteriors are dominated by the CBSH priors, with the exception being $\Omega_{\rm b}h^2 = 0.02233^{+0.00024}_{-0.00023}$, which corresponds to a  $\sim 20\%$ reduction in the size of the CBSH 2-$\sigma$ interval. As in the $w$CDM model, increasing the IGM variance to $\sigm=400$ \pccm weakens this improvement to a few percent, and there is no improvement if the sample size is reduced to $\nfrb=100$. Thus, even under our most optimistic assumptions, we find FRB provide no additional information about the nature of dark energy.

\section{Synergy with 21cm BAO Experiments}
\label{synergies}
Future 21cm Intensity Mapping (IM) experiments designed to measure BAO in the distribution of neutral hydrogen, such as HIRAX and CHIME, are expected to numerous FRBs during the course of their observing runs. Since these FRB detections will essentially come for free (although the redshift will require dedicated observations), we aim determine whether their inclusion in the data analysis might improve the constraint forecasts for the 21cm IM BAO alone. Here we perform a simultaneous MCMC analysis of the FRB1 catalogue with the mock 21cm IM BAO measurement presented in \citep{witzemann}. The mock BAO data is generated for HIRAX, which is a near-future radio interferometer planned to be built in South Africa. It will consist of $1024$ $6$m dishes, covering the frequency range $400$-$800$  MHz, corresponding to a redshifts between $0.8$ and $2.5$. We assume an integration time of $1$ year, and a non-linear cutoff scale at $z=0$ of $k_\mathrm{NL,0} = 0.2$ Mpc${}^{-1}$, which evolves with redshift according to the results from \citep{2003MNRAS.341.1311S}, $k_\mathrm{max} = k_\mathrm{NL,0}(1+z)^{2/(2+n_{\rm s})}$ with the spectral index $n_{\rm s}$. We use these specifications and a slightly adapted version of the publicly available code from \citep{2015ApJ...803...21B} to calculate covariance matrices $\mathbf{C}_\mathrm{BAO}$ for the Hubble rate, $H$, and angular diameter distance, $D_{\rm A}$, in $N=20$ equally spaced frequency bins. We consider correlations between $H$ and $D_{\rm A}$ and assume different bins to be uncorrelated. For the MCMC analysis, the likelihood of a given set of cosmological parameters is then calculated using these measurements together with the FRB1 catalogue, according to
\begin{align}
\ln \mathcal{L}= \ln \mathcal{L}_{\rm BAO} + \ln \mathcal{L}_{\rm FRB} ~,
\end{align}
where
\begin{align}
\ln \mathcal{L}_{\rm BAO} = -\frac{1}{2}\sum_{j=1}^{N} (\nu_j - \mu_j)^\mathrm T \mathbf{C}_\mathrm{BAO}^{-1}(z_j) (\nu_j - \mu_j)
\end{align} 
and $\ln \mathcal{L}_{\rm FRB}$ is given by \er{likelihood}. Further definitions used are $\nu_j = (D_\mathrm A(z_j, \theta) , H(z_j, \theta))$ as well as  $\mu_j = (D_\mathrm A(z_j, \theta_\mathrm{fid}) , H(z_j, \theta_\mathrm{fid}))$. All priors are flat and identical to the ones used in the FRB analysis.

We find that FRBs add little to the constraints coming from 21cm BAO alone --- they only tend to remove some of the non-Gaussian tails in the BAO posteriors. However, they do add an additional parameter into the fitting process, \obhs, which turns out to be the most competitive constraint. We find $\Omega_{\rm b}h^2=0.02235^{+0.00032}_{-0.00032}$, which is comparable to the current CBSH constraint, and entirely independent. This suggest that, when combined with 21cm IM BAO measurements, FRBs may provide an intermediate redshift measure of the cosmological baryon density, independent of high redshift CMB constraints.

\section{Conclusions}
\label{conc}
In this paper we have investigated how future observations of FRBs might help to constrain cosmological parameters. By constructing various mock catalogues of FRB observations, and using MCMC techniques, we have forecast constraints for parameters in the flat \lcdm model, as well as \lcdm with spatial curvature, flat $w$CDM and flat $w_0w_a$CDM. Since $\dmigm \propto \Omega_{\rm b} H_0$, we find  $\Omega_{\rm b}h^2$ and $H_0$ are degenerate, and unconstrained by FRBs observations alone. And as a result, the other cosmological parameters are very weakly constrained, if at all. In all models considered here, the measurement precision on $\Omega_{\rm m}$ is a few tens of percent, when using the most optimistic catalogue with no priors. This is a order of magnitude larger than current constraints coming from CBSH. To determine whether FRBs will improve current constraints, we have included in our FRB analysis realistic priors in the form of the CBSH covariance matrix. With this we showed that \obhs and $\Omega_k$ are the only two parameters that are better constrained when FRBs are included. All dark energy equation of state parameters are poorly constrained by FRBs.

To investigate how sample size and IGM inhomogeneity affect the resulting constraints, we constructed a number of mock catalogues while varying $\nfrb$ and $\sigm$. We find that the inhomogeneity of the IGM poses a serious challenge to the ability of FRBs to improve current constraints. For all model parameterisations that we have considered here, we find that only the most optimistic FRB catalogue gives any appreciable improvement in the current CBSH constraints. For this catalogue we assumed a relatively low DM variance due to the IGM, with $\sigm=200$ \pccm, and a large number of events, with $\nfrb=1000$. Crucially, these events  require followup observations to acquire redshift information, which would require $\sim$100 days of dedicated optical spectroscopic follow-up.   Increasing the IGM inhomogeneity to $\sigm=400$ \pccm, or decreasing the sample size to  $\nfrb=100$ causes the resulting constraints to be dominated by their priors.

Future 21cm IM experiments designed to measure the BAO wiggles in the matter power spectrum will provide independent constraints on cosmological parameters at low/intermediate redshifts. While these observations do not constrain $\Omega_{\rm b}$, they will provide competitive constraints on $H_0$ and $\Omega_{\rm m}$ (within the \lcdm model). Since these experiments are expected to detect many FRBs during the course of their observations, we have investigated combining the BAO constraints with FRB data. We find that this produces a constraint on \obhs comparable to the existing one coming from CBSH observations. Thus, this approach may provide a novel low/intermediate redshift probe of the cosmic baryon density, independent of high redshift CMB data.

The biggest promise of FRB observations seems to be in locating the missing baryons, and not testing concordance or measuring the dark energy equation of state. This may change should one be able to mitigate the effect of IGM variance and the DM contribution from the host galaxy. There are however some caveats. We have assumed that $\figm$ is not evolving with time, and its value is known perfectly.  We have assumed perfect knowledge of $\dmmw$, and that it can be reliably subtracted from $\dmobs$, which is not practical as is know from pulsar observations. Also, we have assumed no error in the redshift of the FRBs. Including these additional sources of uncertainty will weaken any constraints we have obtained here.

\bigskip
We thank Jonathan Sievers and Kavilan Moodley for helpful comments. A. Walters is funded by a grantholder bursary from the National Research Foundation of South Africa (NRF) Competitive Programme for Rated Researchers (Grant Number 91552). A. Weltman gratefully acknowledges financial support from the Department of Science and Technology and South African Research Chairs Initiative of the NRF. The Dunlap Institute is funded through an endowment established by the David Dunlap family and the University of Toronto. B.M.G. acknowledges the support of the Natural Sciences and Engineering Research Council of Canada (NSERC) through grant RGPIN-2015-05948, and of the Canada Research Chairs program. Y.Z.M. acknowledges the support by NRF (no. 105925). A. Witzemann acknowledges support from the South African Square Kilometre Array Project and NRF. Any opinion, finding and conclusion or recommendation expressed in this material is that of the authors and the NRF does not accept any liability in this regard.

\software{emcee \citep{2013PASP..125..306F}, GetDist ( \tt https://github.com/cmbant/getdist)}




\bibliography{frb_refs} 



\end{document}